\documentclass[modern]{aastex61}
\usepackage{graphics}
\usepackage{epsfig}
\usepackage{graphicx}

\def\kms{km~s$^{-1}$}

\usepackage{xcolor}
\definecolor{Green}{rgb}{0,0.5,0}
\definecolor{Blue}{rgb}{0,0,1}
\definecolor{Red}{rgb}{1,0,0}

\received{November 7, 2018}
\revised{\today}

\submitjournal{AJ}

\shorttitle{The Arecibo Pisces-Perseus Supercluster Survey: I.}
\shortauthors{O'Donoghue et al.}

\begin{document}

\title{The Arecibo Pisces-Perseus Supercluster Survey: I. Harvesting ALFALFA}

\correspondingauthor{Aileen A. O'Donoghue}
\email{aodonoghue@stlawu.edu}

\author[0000-0002-0786-7307]{Aileen A. O'Donoghue}
\affil{St. Lawrence University Department of Physics,
23 Romoda Drive,
Canton, NY 13617, USA}

\author{Martha P. Haynes}
\affiliation{Cornell Center for Astrophysics and Planetary Science,
Space Sciences Building, Cornell University, Ithaca, NY 14853, USA}

\author{Rebecca A. Koopmann}
\affiliation{Union College Department of Physics and Astronomy, 
807 Union Street,
Schenectady, NY 12308, USA}

\author{Michael G. Jones}
\affiliation{Instituto de Astrof\'isica de Andaluc\'ia,
CSIC, Glorieta de la Astronom\'ia s/n,
E-18008, Granada, Spain}

\author{Riccardo Giovanelli}
\affiliation{Cornell Center for Astrophysics and Planetary Science,
Space Sciences Building,
Cornell University,
Ithaca, NY 14853, USA}

\author{Thomas J. Balonek}
\affiliation{Colgate University Department of Physics and Astronomy,
13 Oak Drive
Hamilton, NY 13346 USA}

\author{David W. Craig}
\affiliation{West Texas A\&M University Department of Chemistry and Physics,
2403 Russell Long Blvd.,
Canyon, TX 79015 USA}

\author{Gregory L. Hallenbeck}
\affiliation{Washington \& Jefferson College Department of Computing and Information Studies,
60 S. Lincoln St.,
Washington, PA 15301, USA}

\author{G. Lyle Hoffman}
\affiliation{Lafayette College Department of Physics,
Hugel Science Center,
Easton, PA 18045 USA}

\author{David A. Kornreich}
\affiliation{Humboldt State University Department of Physics and Astronomy,
1 Harpst St.
Arcata, CA 95521 USA}

\author{Lukas Leisman}
\affiliation{Valparaiso University Department of Physics and Astronomy,
1610 Campus Drive East,
Valparaiso, IN 46383, USA}

\author{Jeffrey R. Miller}
\affiliation{St. Lawrence University Department of Physics,
23 Romoda Drive,
Canton, NY 13617, USA}

\begin{abstract}
We report a multi-objective campaign of targeted 21 cm ${\rm{H}}{\rm{I}}$ line observations of sources selected from the Arecibo Legacy Fast ALFA (Arecibo L-band Feed Array) survey (ALFALFA) and galaxies identified by their morphological and photometric properties in the Sloan Digital Sky Survey (SDSS). The aims of this program have been (1) to confirm the reality of some ALFALFA sources whose enigmatic nature suggest additional multiwavelength observations; (2) to probe the low signal-to-noise ratio regime, below the ALFALFA reliability limit; and (3) to explore the feasibility of using optical morphology, color and surface brightness to identify gas-rich objects in the region of the Pisces-Perseus Supercluster (PPS) whose ${\rm{H}}{\rm{I}}$ fluxes are below the ALFALFA sensitivity limit at that distance. As expected, the reliability of ALFALFA detections 
depends strongly on the signal-to-noise ratio of the ${\rm{H}}{\rm{I}}$ line signal and its coincidence with a probable stellar counterpart identified by its optical properties, suggestive of on-going star formation. The identification of low mass, star forming populations enables targeted ${\rm{H}}{\rm{I}}$ line observations to detect galaxies with HI line fluxes below the ALFALFA sensitivity limits in fixed local volumes (D $<$ 100 Mpc). The method explored here serves as the basis for extending the sample of gas-bearing objects as part of the on-going Arecibo Pisces-Perseus Supercluster Survey (APPSS).

\end{abstract}


\section{Introduction} \label{sec:intro}

Filamentary structures are predicted to be pervasive throughout the cosmic web, but they have received relatively little individual attention in comparison to other cosmic overdensities such as clusters. The infall of galaxies onto supercluster filaments is a direct prediction of numerical simulations.  The detection of such motions, derived from application of the Tully-Fisher relation (TFR), for example, would provide direct constraints on the mass overdensity along the cosmic web. 

One of the most prominent local extragalactic structures is the Pisces-Perseus Supercluster (PPS), a strong filamentary overdensity first identified as a ``metagalactic cloud'' by \citet{1932Natur.130..132B}. Fortuitously aligned nearly in the plane of the sky \citep[\textit{e.g.}][see Figure \ref{fig:skypps}]{1993AJ....105.1251W} and relatively nearby (c$z\sim$5000 \kms), the main ridge of the PPS connects the clusters Abell 262, 347 and 426 and the rich groups around NGC~7626 (Pegasus I), NGC~383 (Pisces) and NGC~507
\citep[\textit{e.g.}][]{1976PASP...88..388C, 1978IAUS...79..241J, 1981ApJ...243..411G}.
The supercluster was the target of an extensive (at the time) redshift survey by Giovanelli, Haynes and collaborators \citep[e.g.][]{1988lsmu.book...31H}, enabling further studies \citep[\textit{e.g.}][]{1983A&A...121....5C, 1986ApJ...300...77G, 1986ApJ...306L..55H} of the geometry of the supercluster, the morphological segregation of galaxies by galaxy density and the connection between it and our local supercluster, Laniakea \citep{2014Natur.513...71T}. The majority of the redshifts were obtained through ${\rm{H}}{\rm{I}}$ 21~cm line observations with the Arecibo telescope south of its northern declination limit of +36$^\circ$, with others mainly obtained with the former 300-foot telescope at Green Bank, WV. The overall ${\rm{H}}{\rm{I}}$ redshift survey established that the main ridge of the PPS extends at least 50$h^{-1}$ Mpc from the Perseus to Pegasus clusters and confirmed the thin wall of \citet{1978IAUS...79..241J} that is parallel to the sky and 5$h^{-1}$ to 10$h^{-1}$ Mpc deep with voids on both the near and far sides. The foreground void, centered at c$z\sim$3000 \kms, separates the PPS from the Local Supercluster \citep{1986ApJ...306L..55H, 1991ApJ...372L..13W, 2008AJ....135..588S}; its proximity and strong underdensity makes it an excellent laboratory for probes to test the ``Void Phenomenon'' (\citealp{2001ApJ...557..495P, 1994AJ....108..491S}; \citealp[\textit{c.f.} ][]{2009ApJ...691..633T}), the apparent discrepancy between the number of observed galaxies in voids and the number of low mass dark matter halos predicted to reside in them by numerical simulations. 

Studies of large scale structure and motions in the PPS region, bounded here by 22$^{\rm{h}}<$ R.A.$< 03^{\rm{h}}$, 20$^\circ<$ Dec.$<$+45$^\circ$, c$z<$ 9000 \kms ~(see Figure \ref{fig:skypps}), benefit directly from increasing the number of available redshifts and peculiar velocities. Given the large overdensity of PPS, infall onto PPS should be strong. \citet{2001A&A...378..345H} suggest that a galaxy located halfway between the Local Group and PPS should reach an infall velocity of ~500 \kms. The orientation of the PPS at nearly constant distance is optimal, since infall onto the ridge should align with the radial direction. Numerous past TFR studies have attempted detection of the signature of infall onto PPS \citep[\textit{e.g.}][]{1992ApJ...396..453H, 1996ApJ...468L...5D, 1997MNRAS.291..488H, 1997ApJ...475..421E, 1998A&A...340...21T, 1999ApJ...522....1D, 2001A&A...378..345H, 2006PhDT.........2S} but with marginal results. Designed for other purposes, \textit{e.g.}, determining the bulk flow and convergence depth sky-wide, these works generally contain few objects in PPS itself. For example, the extensive SFI++ catalog \citep{2007ApJS..172..599S} only included 350 galaxies in the PPS study of \citet{2006PhDT.........2S}, yielding a signature of inflow onto PPS but only at the 1$\sigma$ level.  Most recently, the 2MASS Tully-Fisher (2MTF) survey \citep{2014MNRAS.445..402H} used ALFALFA ${\rm{H}}{\rm{I}}$ line widths to complement other 21 cm line observations for $\sim$2000 bright inclined spirals to make a measurement of the bulk flow; again, only a small number are in the PPS direction. Likewise, the PPS does not show up prominently in the local universe ``Cosmic Flows'' reconstruction of \citet{2014MNRAS.444..527S} because it lies just outside the volume considered. However, it is evident from that work that the comparison of a well-populated PPS dataset with mock catalogs and constrained simulations will provide a clear measurement of the infall onto PPS and thus a robust estimate of its mass overdensity. 

To make a robust measurement of infall, a substantial increase in peculiar velocity measurements in and near PPS is required. For a fixed volume, the only way to improve on the previous results is to increase the number of peculiar velocity measurements by including galaxies of lower ${\rm{H}}{\rm{I}}$ mass at the PPS distance. Toward that end, we have initiated the Arecibo Pisces-Perseus Supercluster Survey (APPSS) to increase the sampling of peculiar velocities in the PPS direction. Furthermore, because the baryonic mass of star-forming galaxies of low mass is dominated by the ${\rm{H}}{\rm{I}}$ content \citet{2012ApJ...756..113H}, our program aims to extend the previous works by adopting the baryonic version of the TFR: the baryonic Tully-Fisher relation (BTFR) as the scaling law to derive secondary distances and by comparing the observational results with expectations derived from numerical simulations.

A main objective of the APPSS observational program is to increase the sampling of galaxies with redshift and BTFR measurements in the PPS and its foreground and immediate background. This paper presents the results of an initial campaign to obtain ${\rm{H}}{\rm{I}}$ 21 cm line observations of candidate low signal-to-noise ratio (SNR) ${\rm{H}}{\rm{I}}$ signals identified in the ALFALFA extragalactic ${\rm{H}}{\rm{I}}$ survey \citep{2005AJ....130.2613G, 2018ApJ...861...49H} that encompassed much of the relevant PPS volume. Here we review the results of ALFALFA in the PPS region (Section \ref{sec:ALFALFA}) and report on a first set of new observations (Section \ref{sec:observations}) targeting low SNR ALFALFA ${\rm{H}}{\rm{I}}$ candidate detections (Section \ref{sec:results}) using the L-band wide (LBW) receiver system on the Arecibo 305~m antenna. Throughout this work, the assumed cosmology is $H_0 = 70$~\kms Mpc$^{-1}$, $\Omega_m=0.3$, and $\Omega_\Lambda=0.7$.

\section{ALFALFA ${\rm{H}}{\rm{I}}$ line sources in the PPS region} \label{sec:ALFALFA}

The ALFALFA survey \citep{2005AJ....130.2613G, 2018ApJ...861...49H} was a blind 21 cm ${\rm{H}}{\rm{I}}$ line survey of the entire sky visible from Arecibo Observatory outside the zone of avoidance conducted between 2005 and 2012.  It is the most sensitive wide area ${\rm{H}}{\rm{I}}$ survey to date covering 7,000 deg$^2$ and detecting 31,500 extragalactic ${\rm{H}}{\rm{I}}$ line sources out to z $\sim$0.06. A major goal of ALFALFA  wasto make a robust determination of the ${\rm{H}}{\rm{I}}$ mass function, particularly at its faint end, over a cosmologically-significant volume \citep{2005AJ....130.2613G, 2010ApJ...723.1359M, 2018MNRAS.477....2J} and to map the distribution of the population of optically faint, gas-rich systems \citep{2007ApJ...665L..15K}.  The final ALFALFA extragalactic catalog of high-reliability ${\rm{H}}{\rm{I}}$ detections, referred to as $\alpha$.100, has been presented in \citet{2018ApJ...861...49H} and the catalog and spectra are available on the data web site at egg.astro.cornell.edu/alfalfa/data/index.php and will soon be incorporated into the NASA Extragalactic Database (NED; ned.ipac.caltech.edu).

The ``minimum intrusion'' observing technique adopted for ALFALFA delivers consistently high quality spectral data. The published ALFALFA catalog is based on a source identification and measurement scheme intended to deliver a homogeneous and well-understood dataset. Sources were identified with both a matched-filter detection algorithm and visual inspection. The completeness and reliability of the ALFALFA ${\rm{H}}{\rm{I}}$ source catalog is well characterized and understood \citep{2005AJ....130.2613G, 2007AJ....133.2087S, 2011AJ....142..170H}. Within its catalogued compilations, ALFALFA sources are assigned to several distinct categories. Highly reliable extragalactic detections with high SNRs ($\geq$ 6.5), within similar strength in both orthogonal polarizations with no radio frequency interference (RFI) and observed by more than one of the ALFA feeds \citep{2011AJ....142..170H} are very reliable detections and designated ``Code 1''.  Extragalactic sources of lower SNR but which coincide in position and recessional velocity with galaxies of known c$z$, dubbed the ``priors'', are likely to be real and are assigned as ``Code 2''. The Code 1 and 2 sources are included in the ALFALFA extragalactic catalog of \citet{2018ApJ...861...49H}, because of their high degree of reliability. Other lower SNR candidate sources are included in the ALFALFA database; their reliability is discussed in Section 3.

\begin{figure}[t!]
\centering
\includegraphics[width=5.75in]{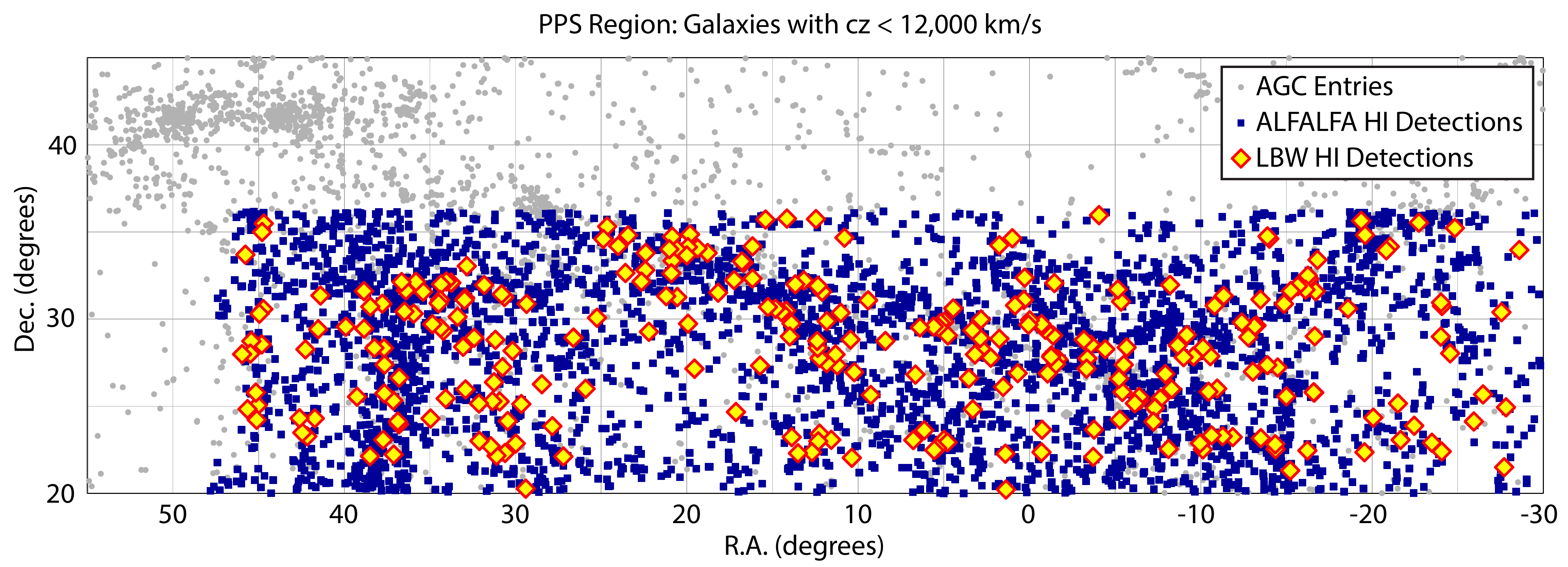}\\
\includegraphics[width=6.0in]{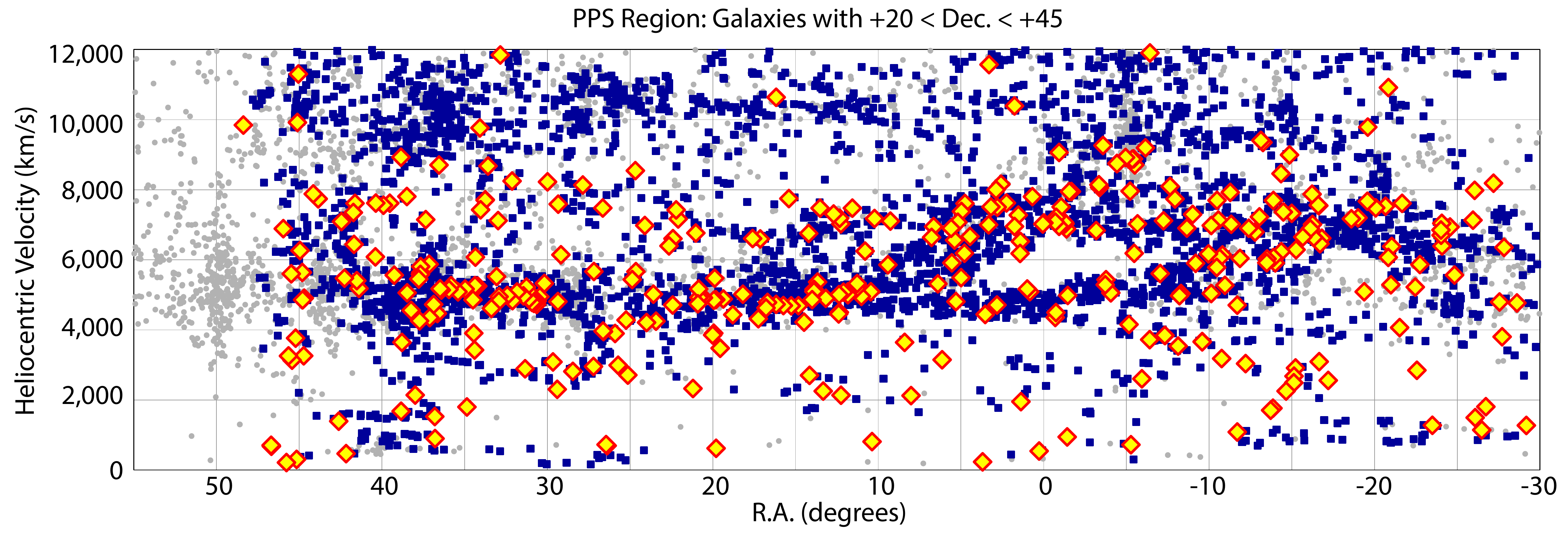}
\vspace{-0.1cm}
\caption{Upper: Sky distribution of galaxies in the PPS region that are included in the AGC compilation (gray), the ALFALFA extragalactic catalog (blue squares) and the targeted observations reported here (yellow diamonds with red borders). In all cases, only objects with c$z$ $<$ 12000 \kms~ are shown. Note the northern boundary of the Arecibo observations at Dec. $\sim$ +36$^{\circ}$. Lower: Distribution of observed heliocentric velocities for galaxies in the same sky region. Symbols are as in the upper panel.}
\label{fig:skypps}
\end{figure}

As part of the early ALFALFA science program, \citet{2008AJ....135..588S} published a catalog of ${\rm{H}}{\rm{I}}$ sources in the Anti-Virgo region at $\delta =$ +27$^\circ$ in a strip covering 135 deg$^2$ within 22$^{\rm{h}} \leq \alpha \leq $ 3$^{\rm{h}}$ and 26$^\circ \leq \delta \leq$ 28$^\circ$. Their catalog included 436 detections with either SNR $>$ 6.5 and 95\% reliability (the Code 1 sources) or classified as ``priors'' with 5.0 $\leq$ SNR $\leq$ 6.5 and an optical counterpart (OC) with a known coincident redshift. In addition, 49 high SNR ${\rm{H}}{\rm{I}}$ signals without OCs with c$z <$ 300 \kms~were identified as likely ${\rm{H}}{\rm{I}}$ high velocity clouds (HVCs) in the Galactic periphery \citep{2013ApJ...768...77A}; they are catalogued separately. Of the detected extragalactic sources, there were 274 (62\%) new ${\rm{H}}{\rm{I}}$ detections and 261 (59\%) newly measured redshifts. Measured properties agree with those presented in the compilation of targeted ${\rm{H}}{\rm{I}}$ line observations by \citet{2005ApJS..160..149S}.  Within cz = 9,000 \kms, known optical galaxies and ALFALFA detections have similar distribution in both space and velocity, strongly peaked on the PPS itself.  Based on the early ALFALFA dataset, \citet{2008AJ....135..588S} showed that a 460 Mpc$^3$ void within the boundaries 22$^{\rm{h}}$ $\leq \alpha \leq$ 2$^{\rm{h}}$ and 1,000 $\leq$ cz $\leq$ 2,500 \kms ~has no ${\rm{H}}{\rm{I}}$ halos of ${M}_{{\rm{H}}{\rm{I}}}$ $\geq 10^8$ M$_\odot$ within it. According to theory \citep[\textit{e.g.}][]{2003MNRAS.344..715G} there should be 38. This result was already an early indication that ALFALFA would not reveal a large population of dark galaxies, a conclusion reinforced as the survey has progressed \citep[e.g.][]{2011AJ....142..170H, 2015AJ....149...72C, 2017ApJ...842..133L}.

The upper panel of Figure \ref{fig:skypps} shows the distribution on the sky of galaxies in the PPS region considered here. Gray circles denote the locations of galaxies with c$z$ $<$ 12,000 \kms ~as included in our redshift database, the Arecibo General Catalog (AGC), a compilation of optical and ${\rm{H}}{\rm{I}}$ galaxy data obtained from the literature plus our own datasets.  Blue symbols mark the ${\rm{H}}{\rm{I}}$ line detections included in the ALFALFA extragalactic catalog \citep{2018ApJ...861...49H} and red ones identify galaxies detected by the targeted ${\rm{H}}{\rm{I}}$ line observations reported in Section \ref{sec:detections}. The absence of blue and red symbols at the eastern edge beyond R.A.~$>$ 3$^{\rm{h}}$ and north of Dec.~$>$ 30$^\circ$ is explained by the boundaries of the ALFALFA survey, and the northern declination limit of the Arecibo telescope. Using the same symbol scheme, the lower panel shows the velocity distribution of galaxies over all detections within the PPS box plotted against their R.A. The PPS is clearly evident as the strong overdensity of points at c$z$ $\sim$5,000 \kms. The general outline of the PPS as presented in Figure 1 of \citet{1993AJ....105.1251W} is strongly reinforced by the newer redshift data contributed by the ${\rm{H}}{\rm{I}}$ line spectra discussed here.

\begin{figure}[t!]
\centering
\plotone{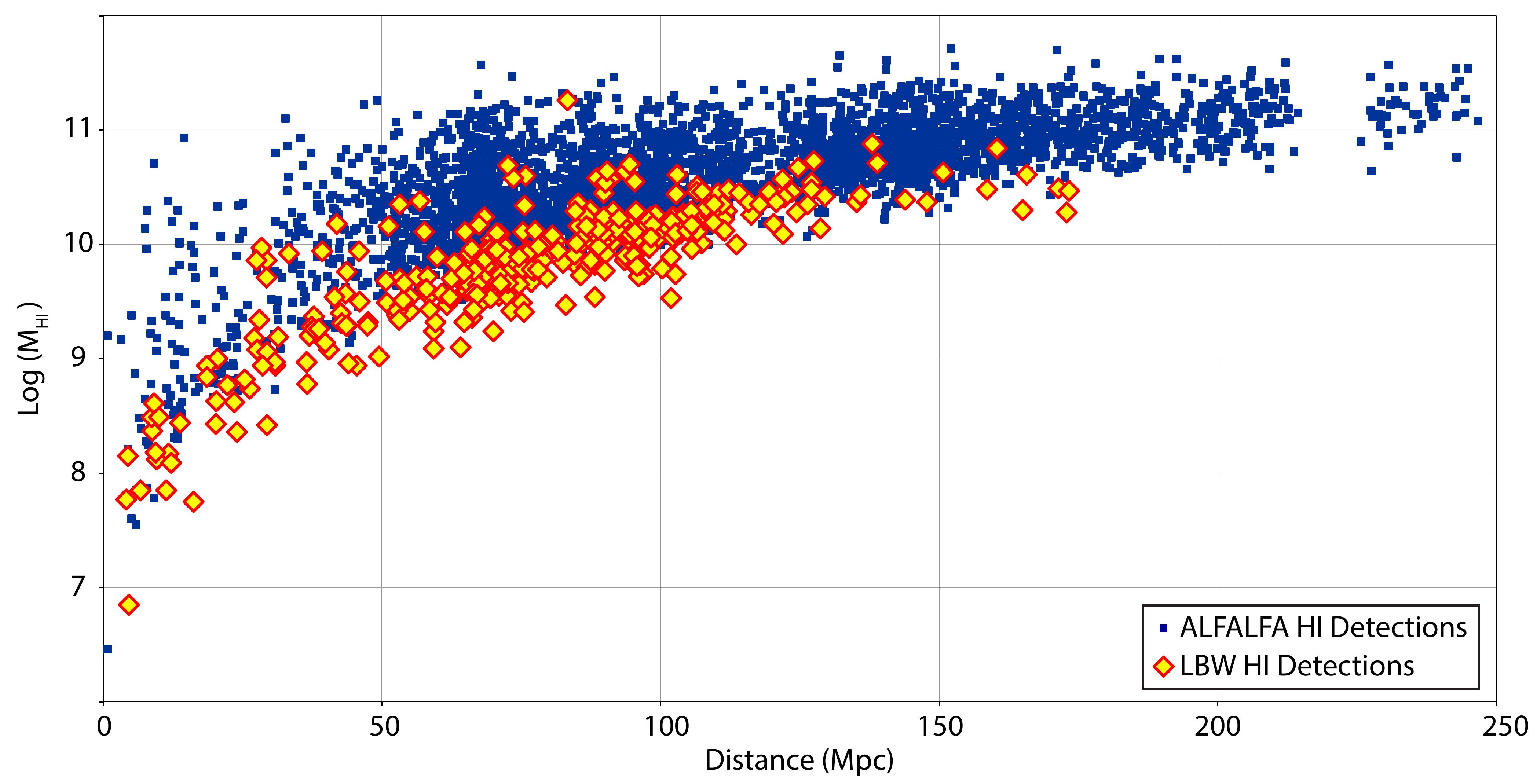}
\vspace{-0.1cm}
\caption{Spanhauer diagram for the ALFALFA galaxies in the PPS region (blue squares) and the ${\rm{H}}{\rm{I}}$ detections made with the LBW observations reported here (yellow diamonds with red borders). As expected, the LBW observations target objects of lower ${\rm{H}}{\rm{I}}$ mass than those detected by ALFALFA. Two points at low ${\rm{H}}{\rm{I}}$ mass/distance have been dropped. The gap in the distribution at about 220 Mpc is an artifact of contamination by RFI associated with the San Juan airport radar.}
\label{fig:span18}
\end{figure}

As illustrated by the blue circles in Figure \ref{fig:skypps}, a large part of the PPS is included in the area covered by the full ALFALFA survey. Within the PPS sub-region covered by the survey (eastern boundary at 3$^{\rm{h}}$10$^{\rm{m}}$ and northern limit of +36$^\circ$10\arcmin , the $\alpha.$100 catalog contains 3,715 detections (high SNR plus priors) with c$z <$ 12,000 \kms, 73\% of which lie at c$z <$ 9,000 \kms. The blue points in Figure \ref{fig:span18} trace the ${\rm{H}}{\rm{I}}$ mass distribution of all the ALFALFA sources in the PPS region as a function of distance out to the bandwidth limit of the survey.  On-going work by this group is aimed at establishing the best approach to application of the BTFR to the ALFALFA sample overall as well as to the galaxies in the PPS region. The ALFALFA dataset overall will make a significant contribution to the studies of velocities and velocity widths in the nearby universe. In the PPS region itself, it provides a large increase in the sampling of galaxies that can be used for peculiar velocity measurements. In the relevant volume, growth of the BTFR peculiar velocity dataset can only be accrued by the extension of samples to lower baryonic masses. An additional blind survey like ALFALFA with deeper integrations would detect more bright galaxies at larger distances than fainter, nearby galaxies. Hence we have been motivated to test other ways of increasing the yield of galaxies in and near the supercluster, with c$z$ $<$ 9,000 \kms. The next section describes a pilot program designed to test ways of achieving that goal.

\section{A Pilot Program to ``Harvest'' ALFALFA in the PPS region} \label{sec:observations}

As noted above, the reliability of ALFALFA  detections is a strong function of the SNR. While the ALFALFA  survey and its reduction pipeline are designed to identify and flag RFI, low level or transient man-made signals may still contaminate the ALFALFA  dataset. As discussed in \citet{2018ApJ...861...49H}, each spectral channel in the final spectrum is assigned a normalized weight reflecting the amount of data contributing to that value that has been excised (flagged). RFI is usually time variable and polarized, so that a suspicious source is one that shows low spectral weight, mismatch in the signal evident in the two polarizations and/or large positional offset of the OC from the ${\rm{H}}{\rm{I}}$ centroid.

High SNR ALFALFA detections have a reliability of 95\% \citep{2018ApJ...861...49H}.  The suspect detections include those with the conditions listed above, those close to the SNR 6.5 lower limit (``ALFALFA High SNR''), and those lacking OCs (``High SNR, no OC''). For low SNR ALFALFA detections, the match-filter detection algorithm is fully expected to become statistically unreliable. However, it was realized that some of the low SNR sources with possible star-forming OCs with a redshift measurement (``Low SNR OC, no z'') might be real, especially at SNRs close to the adopted limit of 6.5. Among the low SNR detections without a suggestive redshift match (``Prior''), two more classes of low SNR candidate ${\rm{H}}{\rm{I}}$ signals are designated internally in the  full ALFALFA  database: (1) those with no identifiable OC (``Nearby, no OC'') and (2) those which coincide with a possible OC but with no reliable optical redshift (``marginal prior''). 

\begin{figure}[t!]
\centering
\epsscale{1.1}
\plotone{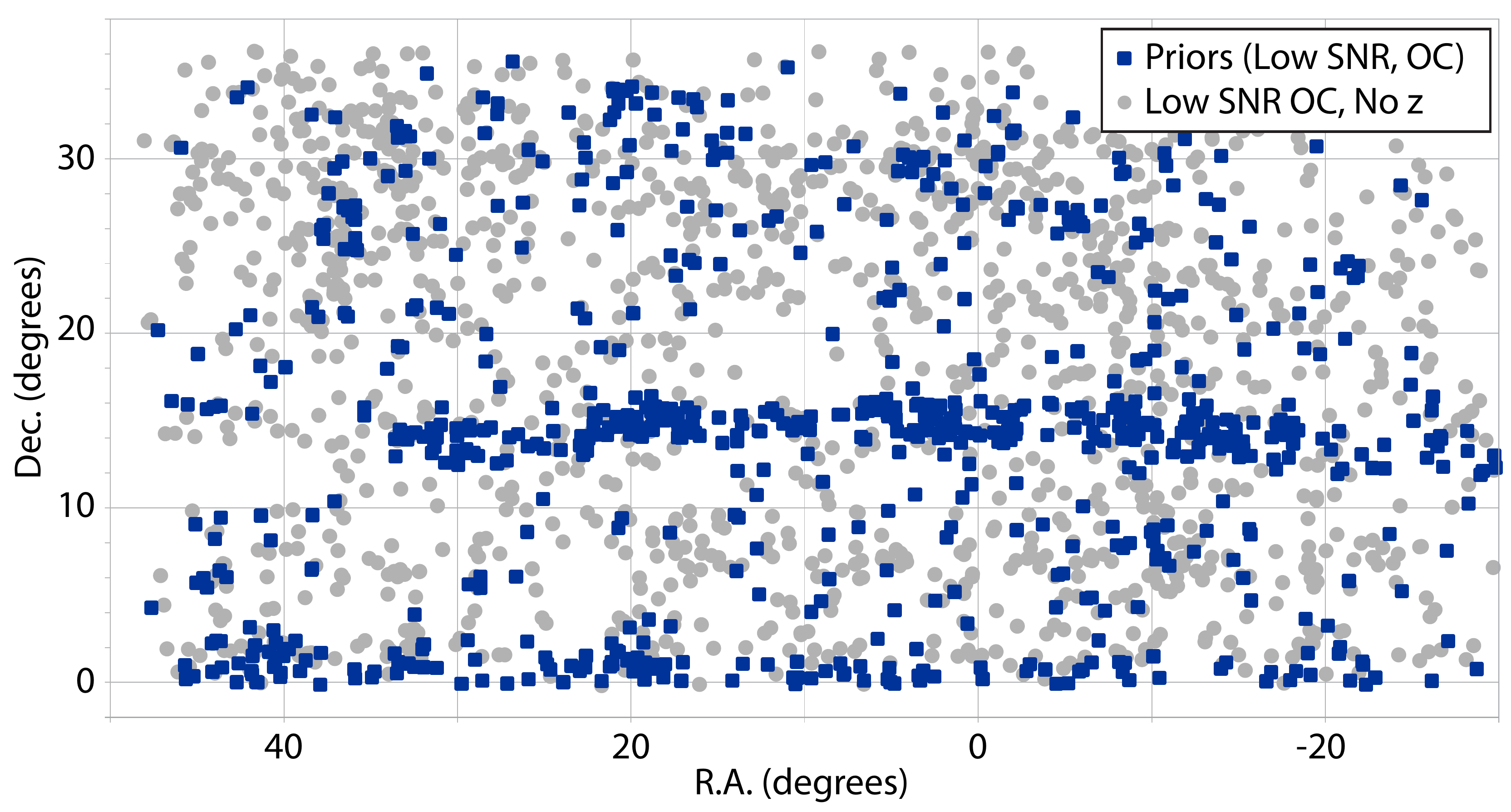}
\vspace{-0.1cm}
\caption{Sky distribution of low SNR sources and ${\rm{H}}{\rm{I}}$ line emission candidates in the ALFALFA southern Galactic hemisphere region. Only the portion of this region covered by $\alpha$.100 and Arecibo is shown (22$^{\rm{h}}<$ R.A.$< 03^{\rm{h}}$, 0$^\circ<$ Dec.$<$+36$^\circ$).  Blue squares indicate the ``priors'' identified with OCs of known and coincident redshift; these sources are included in the ALFALFA catalog \citep{2018ApJ...861...49H}. The gray circles denote the locations of low SNR ${\rm{H}}{\rm{I}}$ line candidates that are possibly associated with OCs whose redshifts are unknown. There are proportionately many more of the latter sources in this part of the ALFALFA footprint than in the northern Galactic hemisphere, mainly because of the lack of overlap of the SDSS spectroscopic survey; the SDSS legacy spectroscopic survey covered only narrow strips along the equator and Dec. = +15$^\circ$, giving rise to the rather narrow blue strips in the figure. The observational program reported here has targeted a selection of the ${\rm{H}}{\rm{I}}$ candidates in the APPSS  region.}
\label{fig:skyfallcodes}
\end{figure}

The LBW observations reported here have been aimed at testing the conditions under which the low SNR ${\rm{H}}{\rm{I}}$ line candidate detections might be of use to expand the ALFALFA data set in the PPS region with modest amounts of telescope time. Toward this end, a series of ``Harvesting ALFALFA'' observing programs were begun in 2012 with two primary goals.  The first aim was to test the reliability of the lower SNR ALFALFA sources with and without OCs.  Relatively isolated ALFALFA sources with no identifiable OCs, ``dark galaxy'' candidates, were a particular aim of the early set of targeted observations.  The ALFALFA sources with OCs of unknown redshift were more numerous in the PPS region (``fall sky'') than the northern cap (``spring sky'').  This is due to the fact that the Sloan Digital Sky Survey (SDSS) that is the primary tool used for identifying OCs has much less spectral coverage in the fall sky than the spring sky \citep[\textit{e.g.}][]{2015ApJS..219...12A}.  If the majority of these sources were detected, this would allow ALFALFA to become an even more valuable resource of PPS galaxy redshifts.

Because of the dearth of SDSS or other optical spectroscopy in the northern hemisphere fall sky, a second goal of our follow-up  program has been to explore how feasible it is to identify ${\rm{H}}{\rm{I}}$-bearing galaxies in and near the PPS using the SDSS photometric data alone (``Optical'').  Since, as illustrated in Figure \ref{fig:skypps}, the PPS traces a filament of galaxies \citep{1993AJ....105.1251W} from $\sim$5,000 \kms~to $\sim$9,000 \kms~with voids in front and beyond, it provides a nearly perfect laboratory for testing such a method.  

Toward these ends, we undertook three modest Arecibo observing programs in the direction of PPS using targeted observations with the single-pixel, higher gain LBW receiver. It should be noted that during this same time frame, the ALFALFA dataset was being processed so that its completeness continued to evolve; ALFALFA sources in the observational program reported here were culled from the ALFALFA database compiled before 2016.  For this reason, the detection statistics presented here are meant to be illustrative but are not the result of a statistically robust experiment.

\subsection{Target Selection} \label{subsec:targetselection}

The first observing program undertaken as part of this effort (A2707) aimed to perform follow-up observations on two categories of ALFALFA detections.  A first group of targets consisted of high signal-to-noise objects (SNR $>$ 6.5, Code 1 in $\alpha$.100) deemed marginal due to being close to the SNR boundary of 6.5 and/or lacking OCs (``ALFALFA High SNR'' and ``High SNR, no OC'').  Even high SNR detections in ALFALFA can be spurious \citep{2007AJ....133.2087S}.  If these proved to be real cosmic objects, they could be dark galaxies, tidal tails \citep[\textit{e.g.,}][]{2008ApJ...682L..85K, 2007ApJ...665L..19H}, OH megamasers at 0.16 $<$ z $<$ 0.25 as described by \citet{2002ApJ...572..810D} or HVC’s \citep{2013ApJ...768...77A}.  The second selection of ALFALFA sources consisted of
low SNR sources ($<$ 6.5) lacking OCs with low recessional velocities (c$z <$ 1,000 \kms) with narrow ${\rm{H}}{\rm{I}}$ line widths, suggesting they might be local very gas-rich dwarf galaxies difficult to detect optically.  

The second ALFALFA follow-up observing program (A2811) was primarily to confirm low SNR detections with possible OCs having no available redshift measurements or with velocities in conflict with candidate ALFALFA ${\rm{H}}{\rm{I}}$ signals (``Low SNR OC, no z'').  These objects are not reported in the published ALFALFA catalogs because they do not meet the SNR threshold, nor benefit from coincidence in known redshift with an OC. A few marginal high SNR objects from ALFALFA were also included (``ALFALFA High SNR'').  In addition, some targets were selected visually from the SDSS near clusters with appearances similar to OCs (``Optical'') in the PPS region contained in the available ALFALFA catalog (roughly 70\% of the final catalog).

The third ALFALFA follow-up program (A2899) aimed at confirming more low SNR sources with optical counterparts of unknown or different redshift in the PPS region (``Low SNR OC, no z'') plus additional galaxies selected from SDSS on the basis of optical color and surface brightness (``Optical'').  These targets were chosen in a variety of ways. Initially, Haynes and others, having used SDSS for identifying optical counterparts for the thousands of ALFALFA galaxies and thus being very familiar with their appearances, identified galaxies in SDSS with morphological appearances similar to known OCs. Basically, galaxies with blue colors, low surface brightness, and moderate angular extent were identified and examined in the public SDSS imaging database (in 2014). Objects that showed the patchy, blue morphologies typical of low mass, star-forming galaxies were then targeted for LBW observations. 

The different types of targets described above are summarized in Table \ref{tab:CategoryRates}.

\begin{figure}[ht!]
\plotone{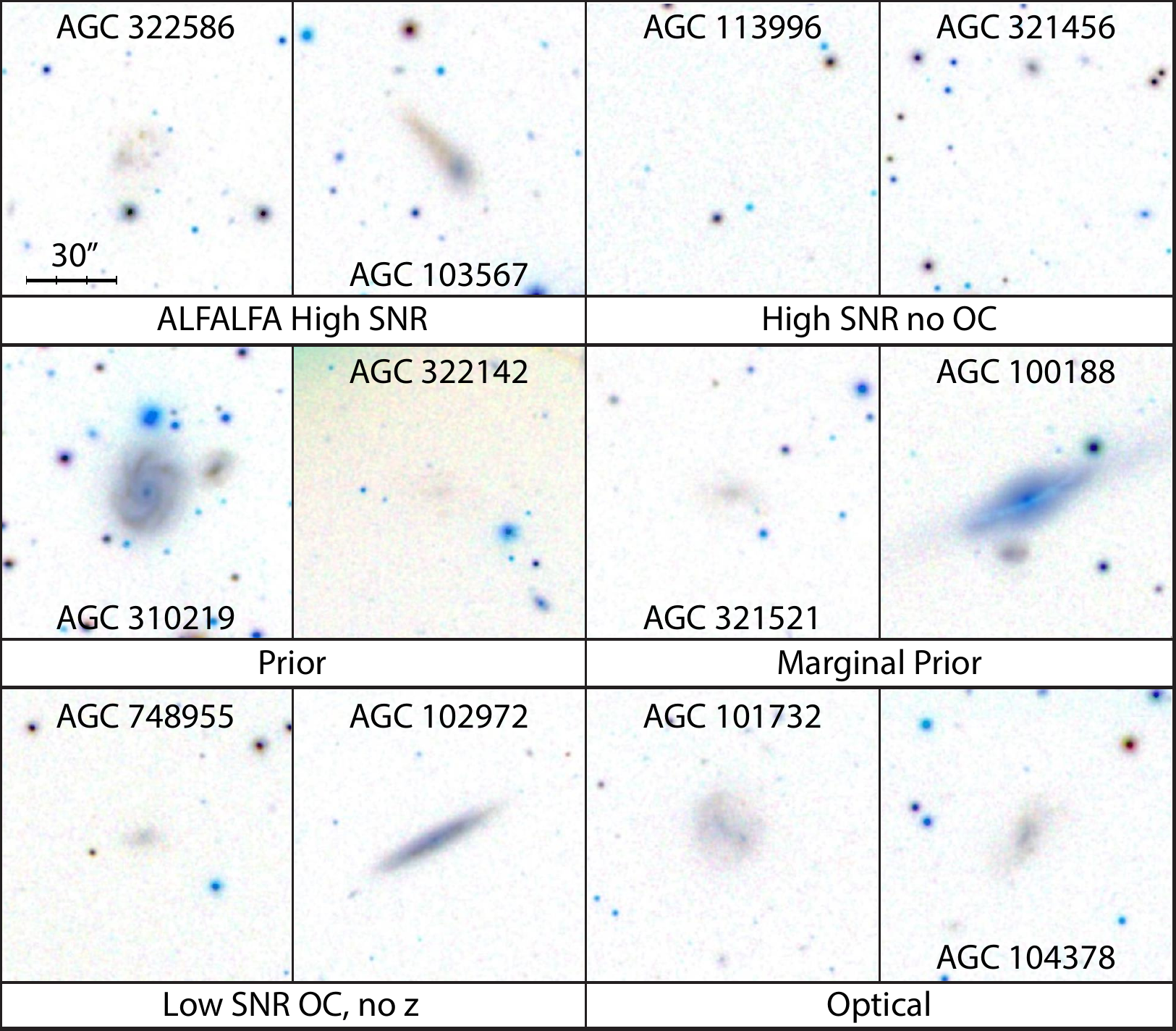}
\caption{SDSS DR14 images of a selection of galaxies detected in these LBW observations.  These were chosen to display the variety of galaxies within each category while revealing the general appearance of galaxies in the LBW flux and distance ranges. Each image is centered on the LBW target coordinates, either the location of the OC or the ${\rm{H}}{\rm{I}}$ centroid derived from the ALFALFA gridded data where no OC has been identified.
\label{fig:GalaxiesNoGrids}}
\end{figure}

An array of sample galaxies in various selection categories detected in these observations is presented in Figure \ref{fig:GalaxiesNoGrids} to illustrate both the variety and similarities of the LBW targets. A limitation of the pilot program presented here is that the frequency range searched for ${\rm{H}}{\rm{I}}$ line emission was limited to 25 MHz usually centered on 5,000 \kms~or on the velocity of a nearby galaxy of known redshift (this corresponds to a 5,450 \kms~span in recessional velocities centered on 5,000 \kms, still within the PPS and foreground void). This was a preliminary attempt to develop photometric selection criteria that have been later applied to the APPSS program to search for ${\rm{H}}{\rm{I}}$ signals over a much wider frequency range and will be described in later papers.

It is important to emphasize that the multi-purpose nature of these observing programs and the preliminary nature of some of the ALFALFA data at the time render the resultant dataset rather inhomogenous. Thus while robust statistical tests are beyond the scope of this work, confirmation of unusual signals, tests of methods of identifying low luminosity ${\rm{H}}{\rm{I}}$-bearing galaxies at $\sim$70 Mpc and further analysis of ALFALFA completeness and instrumental effects all contribute to setting the stage for the APPSS.

\subsection{Observing Methods} \label{subsec:methods}

All observations were conducted with the single-beam LBW receiver system. A 1370-1470 MHz filter was used to limit the impact of human-generated RFI. At zenith, the gain of LBW is about 10.5 K~Jy$^{-1}$ and the system temperature is typically 25~K.

The LBW observations used a basic pointed total-power (TP) position-switched (``ON-OFF'') mode of varying bandwidth (see below) in the 1340-1430 MHz ALFALFA frequency range.  In TP position switching, a source (the ``ON'' scan) is tracked for some time period (usually three to five minutes), then the telescope is slewed back to the original altitude and azimuth location, and tracking of that position (the ``OFF'' scan) occurs for the same length of time.  A diode of known thermal resistance is fired at the end of each ON/OFF source pair to provide calibration. Application of the TP mode assumes that the OFF scan does not contain a source of ${\rm{H}}{\rm{I}}$ line emission at the same frequency as the target galaxy.  The OFF observation allows for bandpass subtraction since the effects of the instrument and the sky should be identical to the ON observation, but without any source emission.  For the observations reported here, we almost always used a 3-minute ON-OFF switching.  This observing scheme was designed to allow detection of ${\rm{H}}{\rm{I}}$ masses greater than $10^7$ M$_\odot$  for $\langle$SNR$\rangle$ $\sim$5.5, an average velocity width (at half the peak flux) of $\langle W_{50}\rangle$ $\sim$100 \kms~ and a smoothing factor of 1/2 \citep[\textit{e.g.}][\S 4.1]{2005AJ....130.2613G}.

\begin{figure}[ht!]
\plotone{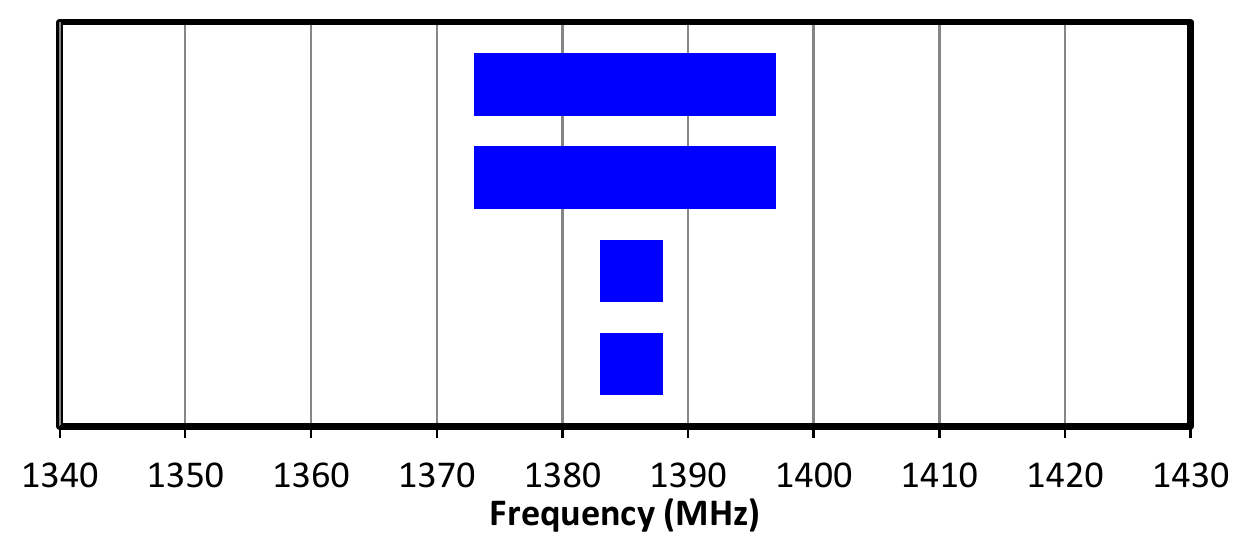}
\caption{Schematic diagram showing the hybrid/dual bandwidth Doppler tracking mode frequency ranges (blue rectangles) for a source with a heliocentric velocity of 7,600 \kms~ centered at 1385 MHz. This configuration of the four quadrants of the interim correlator deliver two 2048-channel spectra per polarization of lower and higher spectral resolution covering 25 and 6.25 MHz respecively.
\label{fig:DualBandwidth}}
\end{figure}

For sources of known redshift, the adopted spectrometer was the ``interim correlator'' used in a hybrid/dual-bandwidth, Doppler-tracking mode. Because the redshift was already known, we acquired two spectra of 2048 channels each but with different spectral resolution (bandwidth) centered on the frequency corresponding to the target's recessional velocity (rounded to the nearest 100 \kms) for each of the two orthogonal polarizations, using 9-level sampling. The lower resolution spectra covered 25 MHz with a resolution per channel of 12.2 kHz or 2.5 \kms~at c$z \sim$0 \kms, while the higher resolution ones extended over the inner 6.25 MHz. Figure \ref{fig:DualBandwidth} illustrates the set-up for a source with a heliocentric velocity of 7,600 \kms~centered at 1385 MHz.

There was no particular noise advantage to configuring the correlator this way, but since the capability of recording separate spectra of varying resolution was available, we decided to obtain spectra at the two different resolutions so as to characterize more accurately both wide and narrow emission lines. Because of the importance of understanding uncertainties introduced by spectral resolution, we observed a small set of galaxies with narrow ($<$ 60/kms) profile widths which are being used to investigate resolution and smoothing effects.

\begin{figure}[ht!]
\plotone{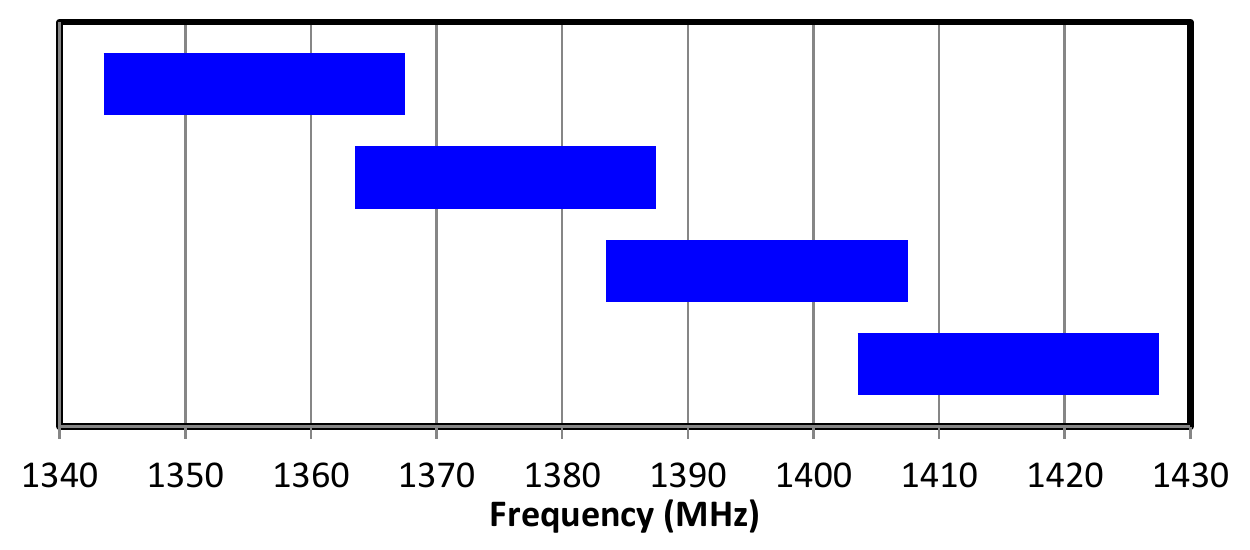}
\caption{Illustration of the four staggered frequency ranges, each of 25 MHz, used in the WAPP search mode for a source of unknown redshift. The frequency boards range from 1343.4058 MHz to 1427.4058 MHz with each board offset from its neighbors by 20 MHz.
\label{fig:WAPPSearch}}
\end{figure}

On two of our last nights of observing, we observed sources of unknown redshift in the third observing program (A2899)
using the Wideband Arecibo Pulsar Processors (WAPPs) in ``search'' mode as a test.  The WAPP search mode was set up to record 8 spectra as polarization pairs distributed over 4 staggered frequency ranges of 25 MHz each.  As shown in Figure \ref{fig:WAPPSearch}, we utilize four overlapping frequency boards ranging from 1343.4058 MHz to 1427.4058 MHz.  Each board is offset from its neighbor(s) by 20 MHz.  This setup insures that any source visible at a frequency near the edge of a board will be seen in both boards and avoids a loss of sensitivity near the edges dude to the bandpass shape. 

The search mode frequency range includes the frequencies of the main beam of the 1350 MHz San Juan airport FAA radar and common harmonics seen at 1380 MHz and 1410 MHz.  They also include the 1381 MHz 30-180 second bursts of the NUclear DETonation (NUDET) system aboard the Global Positioning System (GPS) satellites.  The short NUDET bursts can be removed during data processing, but the airport radar interference blinds us in those frequency ranges. Plots of the frequency channel ``occupancy'' by RFI for various subsets of the ALFALFA data acquisition are shown in Figure 2 of \citet{2018ApJ...861...49H}.

Most sources were observed for a single TP ON-OFF pair but sometimes, mainly due to corruption by RFI, a target was reobserved at a later time. In such cases, a final TP difference spectrum was constructed from the accumulation and averaging of the individual pairs. The combined TP difference spectrum of each observation was then visually inspected to determine whether a source was detected. Baselines were fitted with polynomials of order 0 through 9 with the fit determined by visual inspection and RMS values, selecting the lowest order good fit. Hanning smoothing was applied to all spectra.  Following procedures described in \citet{2005ApJS..160..149S} and \citet{2018ApJ...861...49H}, the ${\rm{H}}{\rm{I}}$ line fluxes $S_{obs} = \int S(v)dv$  in units of mJy-\kms~were calculated by summing the values of the flux in each channel over a frequency span selected by the user. ${\rm{H}}{\rm{I}}$ line velocities and velocity widths were measured from Gaussian fits for single-peaked profiles or, for two-horned ones, by measuring the width across polynomials (usually lines) fit to each of the two sides of the profile between 15\% and 85\% of the peaks on each side. Velocities and velocity widths were measured at fixed levels of 20\% and 50\% of the peak. The rms noise was measured over the signal-free portion of the spectrum.  The SNR was calculated as it was for the ALFALFA data as set forth in Equation 4 of \citet{2018ApJ...861...49H}. 

All the spectra were recorded with 9 level sampling to maximize efficiency ($\sim$96\% of the SNR of analog correlation). Although there is a substantial range due to the non-uniformity of the observing strategy, typical values of the rms noise were about 1.5 mJy at 10 \kms, compared to about 2.5 mJy for the ALFALFA survey itself.

\section{Results} \label{sec:results}

This section gives a summary of the results of the several pilot observing programs and presents the ${\rm{H}}{\rm{I}}$ line parameters for detections, notably those not included in the ALFALFA catalog \citep{2018ApJ...861...49H}.

\subsection{Detection statistics} \label{sec:detections}
Table \ref{tab:CategoryRates} summarizes the detection statistics of the objects in the various selection categories. The high SNR (Code 1) ALFALFA sources remain quite reliable with a detection rates of 100\% for those with OCs and 88.2\% for those lacking an OC. The reliability of high SNR sources has been estimated to be $\sim$95\% \citep{2007AJ....133.2087S}, but it should be noted that the targets observed here were specifically selected based on an increased likelihood of their being spurious (no OC; possibly RFI-contaminated, etc).  Most of the high SNR Code 1 ALFALFA sources lacking optical counterparts are tidal tails. {\citet{2011AJ....142..170H} \S4.3 noted that fewer than 2\% of the high SNR detections were not assigned a possible OC.  Of those, 3/4 are located in fields with objects of similar redshift with a high probability of being part of an extended ${\rm{H}}{\rm{I}}$ distribution or tidal debris field.  Furthermore, \citet{2016MNRAS.463.1692L} showed just how extensive tidal features can be. Some of the high SNR objects without OCs turn out to be isolated very low surface brightness (LSB) galaxies. There have been several campaigns to follow-up on individual optically dark galaxies.  \citet{2016MNRAS.463.1692L} have shown that a few of them do have associated stellar populations upon further investigation or deeper imaging.  For example, the single ${\rm{H}}{\rm{II}}$ region in the nearby dwarf Leo P shows up in the SDSS \citep{2013AJ....146...15G} and the very low surface brightness galaxy HI1232+20 (Coma P) is visible in a GALEX image \citep{2015ApJ...801...96J}.  A sample of 115 very LSB but extended ${\rm{H}}{\rm{I}}$-rich ALFALFA galaxies was studied by \citet{2017ApJ...842..133L} and further works on individual dark and (``almost'') dark galaxies are published or on-going. The low SNR sources detected by ALFALFA with optical counterparts having similar redshifts, the priors, are very reliable since all 9 targets were detected.  Of the priors deemed untrustworthy due to having very low SNRs or low reliability weight due to RFI or other issues, we detected 92.3\% of the targets.  The low SNR sources detected by ALFALFA but having optical counterparts of different or unknown optical redshifts were detected at a rate of 70.2\%.  Thus the presence of a likely OC in the ALFALFA data increases the reliability of the detection even if the OC redshift is unknown or the ALFALFA data is somewhat suspect. 

The clearest unambiguous result was that the sample of very nearby (cz $<$ 1,000 \kms) low SNR ALFALFA detections lacking any optical counterpart: all proved to be false detections.  None of the 84 targets with low SNRs and lacking an OC or any association with a known tidal stream were detected by LBW. 

The detection rate for targets selected by members of our team experienced with determining optical counterparts of ALFALFA sources and by using SQL queries with photometric parameters was 62.5\%.  This gives support to our choice to select the APPSS targets from SDSS using SQL queries.  ALFALFA has taught us that nearly all diffuse star forming, blue, low-mass galaxies contain atomic gas. 

\begin{deluxetable}{lccc}[ht]
\tablecaption{Detection Rates by Category \label{tab:CategoryRates}}
\tablecolumns{4}
\tablenum{1}
\tablehead{
\colhead{Category} &
\colhead{Detections} &
\colhead{Non-Detections} &
\colhead{Detection Rate (\%)}
}
\startdata
ALFALFA High SNR\tablenotemark{a} & 43 & 0 & 100.0 \\
High SNR no OC\tablenotemark{b} & 30 & 4 & 88.2 \\
Prior\tablenotemark{c} & 9 & 0 & 100.0 \\
Marginal Prior\tablenotemark{d} & 12 & 1 & 92.3 \\
Low SNR OC, no z\tablenotemark{e} & 207 & 88 & 70.2 \\
Nearby, no OC\tablenotemark{f} & 0 & 84 & 0 \\
Optical\tablenotemark{g} & 35 & 21 & 62.5 \\
\enddata
\tablenotetext{a}{ High SNR ALFALFA sources with low spectral weight, mismatch in polarizations, large positional offset of ${\rm{H}}{\rm{I}}$ centroid or near the SNR = 6.5 limit}
\tablenotetext{b}{ High SNR ALFALFA sources with no OC.}
\tablenotetext{c}{ Low SNR ALFALFA sources with an OC of known and coincident redshift.}
\tablenotetext{d}{ Priors defined in note c, but deemed untrustworthy due to polarization mismatch, low weight or suspected contamination by RFI }
\tablenotetext{e}{ Low SNR ALFALFA candidate sources with an OC of unknown redshift}
\tablenotetext{f}{ Low SNR ALFALFA candidate sources of low velocity (cz $<$ 1000 \kms), narrow width and lacking an OC; none were confirmed}
\tablenotetext{g}{ Targets chosen from SDSS visually or by SQL Query}
\end{deluxetable}	
 
\subsection{${Properties \ of \ \rm{H}}{\rm{I}}$ Line Detections} \label{subsec:HIdata}

Table \ref{tab:HIData} presents, in order of Right Ascension, the ${\rm{H}}{\rm{I}}$ parameters derived from the LBW spectra.  The first column is the ID number from the AGC, the informal private redshift database of MPH and RG.  Columns 2 and 3 are the J2000 coordinates of the position targeted for the observations; this position is either the location of the OC or, where none was identified, the ${\rm{H}}{\rm{I}}$ centroid derived from the ALFALFA gridded data. It should be noted that the uncertainty in the ${\rm{H}}{\rm{I}}$ positions is much greater than that of typical OC positions and depends on the SNR; the median offset of the two for the ALFALFA survey is $\sim$18\arcsec~on average but is likely to be substantially larger for the low SNR candidate detections reported here. For that reason, the positions of the suspected OCs are targeted. Column 4 gives the source of the target as described in Table \ref{tab:CategoryRates}.  Blank lines are left between categories to distinguish the different source catalogs of targets.  Columns 5 -- 7 are the measured quantities of heliocentric velocity as measured at the center of the ${\rm{H}}{\rm{I}}$ line at the 50\% level, the velocity width of the ${\rm{H}}{\rm{I}}$ line at that level (corrected for channel broadening) and the ${\rm{H}}{\rm{I}}$ line flux as explained above.  Column 8 is the distance determined from the ALFALFA flow model \citep{Masters2005} that includes the effects of the multiple attractors on objects with cz $\leq$ 6,000 \kms.  Beyond 6,000 \kms, the distances are calculated from the Hubble Flow using $D = H_0/v$ assuming $H_0$ = 70 \kms/Mpc.  Column 9 is the log of the ${\rm{H}}{\rm{I}}$ mass found from $M_{{\rm{H}}{\rm{I}}} = 2.36 \times 10^5 D^2 \int S(v)dv$ where the integral was the integrated flux given in column 7 and the gas is assumed to be optically thin. The full table is available electronically as a machine-readable file. 

\begin{deluxetable}{clllRRRrc}
\tablecaption{Properties of LBW Detections \label{tab:HIData}}
\tablecolumns{9}
\tablenum{2}
\tabletypesize{\footnotesize}
\tablehead{
\multicolumn{1}{c}{AGC ID} &
\multicolumn{1}{c}{R.A.} &
\multicolumn{1}{c}{Dec.} &
\multicolumn{1}{c}{Source\tablenotemark{a}} &
\multicolumn{1}{c}{$v_{helio}$} &
\multicolumn{1}{c}{$W_{50}$}  &
\multicolumn{1}{c}{$\int Sdv$} &
\multicolumn{1}{c}{Dist\tablenotemark{b}} &
\multicolumn{1}{c}{log$(M_{HI})$} \\
\multicolumn{1}{c}{} &
\multicolumn{1}{c}{J2000} &
\multicolumn{1}{c}{J2000} &
\multicolumn{1}{c}{Category} &
\multicolumn{1}{c}{\kms} &
\multicolumn{1}{c}{\kms}  &
\multicolumn{1}{c}{J km/s} &
\multicolumn{1}{c}{Mpc} &
\multicolumn{1}{c}{log$(M_\odot)$}
}
\colnumbers
\startdata
103567 & 0.277917 & 32.37833 & ALFALFA high SNR & 542.6 \pm 0.2 & 32.4 \pm 0.5 & 1.27 \pm 0.016 & 8.81 & 7.37\\
101808 & 1.371667 & 20.22667 & ALFALFA high SNR & 1951.6 \pm 1 & 59.1 \pm 2.3 & 0.67 \pm 0.022 & 27.60 & 8.08\\
103722 & 3.691667 & 10.81306 & ALFALFA high SNR & 234.8 \pm 0.1 & 22.6 \pm 0.2 & 1.48 \pm 0.014 & 4.11 & 6.77\\
\hline
102983 & 8.406667 & 28.73917 & High SNR no OC & 3642.6 \pm 4 & 130.1 \pm 5.6 & 0.8 \pm 0.049 & 50.65 & 8.69\\
102879 & 12.35042 & 31.90611 & High SNR no OC & 4503.4 \pm 0.6 & 41.4 \pm 1.5 & 0.74 \pm 0.023 & 65.84 & 8.88\\
102885 & 14.20083 & 30.47583 & High SNR no OC & 6749.7 \pm 1.1 & 43.9 \pm 2.6 & 0.68 \pm 0.035 & 92.04 & 9.13\\
\hline
110158 & 18.79083 & 33.79028 & prior & 4429.5 \pm 3.7 & 177.3 \pm 5.2 & 0.83 \pm 0.055 & 43.50 & 8.57\\
115416 & 19.93917 & 34.13277 & prior & 3905.5 \pm 3.4 & 126 \pm 4.8 & 0.58 \pm 0.043 & 68.69 & 8.81\\
115573 & 21.02542 & 34.0125 & prior & 6762.6 \pm 7 & 158.7 \pm 10 & 0.85 \pm 0.049 & 92.60 & 9.24\\
\hline
100188 & 5.39 & 22.59306 & marginal prior & 6561.5 \pm 7.4 & 325.7 \pm 10.5 & 0.51 \pm 0.06 & 88.87 & 8.98\\
100196 & 5.55625 & 22.48278 & marginal prior & 5910.2 \pm 1.9 & 57.6 \pm 4.8 & 0.34 \pm 0.024 & 79.59 & 8.71\\
111963 & 20.62125 & 34.04 & marginal prior & 4913.3 \pm 3 & 228.4 \pm 4.3 & 0.74 \pm 0.054 & 68.05 & 8.91\\
\hline
102946 & 0.055417 & 29.67695 & low SNR OC no z & 7021.6 \pm 2.4 & 91.6 \pm 5.8 & 0.61 \pm 0.034 & 95.53 & 9.12\\
102597 & 0.682917 & 26.89694 & low SNR OC no z & 7801.7 \pm 0.1 & 316.3 \pm 20.7 & 1.2 \pm 0.061 & 106.60 & 9.51\\
100013 & 0.84125 & 30.78167 & low SNR OC no z & 5086.3 \pm 2.1 & 206.6 \pm 2.9 & 0.78 \pm 0.049 & 72.05 & 8.98\\
\hline
104289 & 0.320417 & 31.10194 & optical & 12461.7 \pm 3.6 & 209.3 \pm 5.1 & 0.42 \pm 0.042 & 173.30 & 9.47\\
104295 & 0.9975 & 34.63834 & optical & 5167 \pm 1 & 25.8 \pm 2.3 & 0.21 \pm 0.016 & 73.23 & 8.42\\
101201 & 1.407917 & 22.28222 & optical & 6524.3 \pm 8.6 & 161.3 \pm 21.5 & 0.36 \pm 0.041 & 88.25 & 8.82\\
\hline
104471 & 12.46042 & 35.73139 & calibrator/test & 12398 \pm 2.1 & 51.4 \pm 5 & 0.27 \pm 0.023 & 172.87 & 9.28\\
115352 & 15.40417 & 35.70945 & calibrator/test & 7748.3 \pm 2.6 & 149.7 \pm 3.7 & 1.09 \pm 0.063 & 106.53 & 9.47\\
122900 & 42.61375 & 24.30944 & calibrator/test & 1392.1 \pm 0.1 & 28.3 \pm 0.6 & 0.86 \pm 0.017 & 18.56 & 7.84\\
\hline
103714 & 3.530417 & 26.61083 & test/other & 4441.4 \pm 3.3 & 117.4 \pm 4.6 & 0.36 \pm 0.034 & 62.51 & 8.52\\
114831 & 19.55125 & 27.14694 & test/other & 3476.6 \pm 3.7 & 66.8 \pm 5.2 & 0.39 \pm 0.024 & 47.45 & 8.32\\
114833 & 26.61167 & 28.93583 & test/other & 3951.2 \pm 1.8 & 152.3 \pm 2.5 & 0.48 \pm 0.041 & 53.79 & 8.52\\
\enddata
\tablenotetext{a}{The categories of targets are detailed in section \ref{subsec:targetselection} and the detection rates presented in Table \ref{tab:CategoryRates}}
\tablenotetext{b}{Distances from the ALFALFA flow model \citep[\textit{see}][]{2011AJ....142..170H}}
\tablecomments{ This table is available in its entirety in machine-readable form.}
\end{deluxetable}

\section{Discussion} \label{sec:Discussion}

The main objectives of this program have been to test the reliability of sources in the ALFALFA catalog, particularly of those with low SNR, and to explore the viability of using complementary optical photometry from SDSS to identify low mass star-forming galaxies that lie in the volume in and around the PPS and are likely to be detected in the ${\rm{H}}{\rm{I}}$ line with modest targeted observations. Due to its heterogeneous nature and the preliminary nature of the ALFALFA dataset in some regions when the observations were undertaken, the statistical completeness of the program is not robust. However, we use the observations to examine some lessons learned.

\subsection{ALFALFA Selected Targets} \label{subsec:ALFALFAtargets}

Given the very large size of the ALFALFA dataset, \citet{2007AJ....133.2087S} predicted that some bright ALFALFA “detections” would not be real purely on the basis of statistical probabilities. Although the drift scan data were flagged carefully and both polarizations are retained for examination, the unpredictable and insidious nature of RFI induces further likelihood of spurious detections. The identification with an OC is a strong confirming indicator so that the few sources without OCs (candidate ``dark'' galaxies) are the most suspect. Optically dark sources could be associated with tidal debris or very extended ${\rm{H}}{\rm{I}}$ disks but some could also be bona fide galaxies with undetectable starlight or interloping OH megamasers at 1.6 $< z <$ 0.22. Because of the scientific importance of understanding the nature of any optically-dark sources, high SNR ALFALFA detections without OCs were reobserved with LBW to confirm their reality as ${\rm{H}}{\rm{I}}$ signals. Thirty-four of these sources were observed, and only four were not confirmed. Those four are AGC 321502 (SNR = 6.9), AGC 333605 (SNR = 7.9), AGC 749115 (SNR = 7.0) and AGC 333631 (SNR = 6.5), all of which were marginally above the ALFALFA SNR threshold. In addition to these (almost) dark high SNR ALFALFA sources, numerous other selected low SNR sources were observed in order to obtain more robust results based on higher SNR spectra.

The targeting of a set of low SNR signal candidates without OCs but likely to be in the very local universe was motivated by the possibility that narrow but weak ${\rm{H}}{\rm{I}}$ line signals might be associated with very low ${\rm{H}}{\rm{I}}$ mass galaxies with only minimal stellar populations. Not unexpectedly, these low velocity (c$z <$ 1,000 \kms) low SNR ALFALFA candidates without optical counterparts are extremely unlikely to be real; in fact, none of the 84 such targets were detected by the deeper LBW observations.  Thus in addition to low ${\rm{H}}{\rm{I}}$ statistical significance, the lack of an optical counterpart in a low SNR source is a strong indication that a ${\rm{H}}{\rm{I}}$ signal candidate is not real.

In contrast to the small number of optically-dark high SNR sources without probable OCs, coincidence of a low SNR ${\rm{H}}{\rm{I}}$ signal with a possible OC increases the chance of the reality of such signals. 

All the published ALFALFA catalogs have included a class of low SNR ${\rm{H}}{\rm{I}}$ sources, the ``priors'', which coincide in position and redshift with likely OCs but would otherwise not meet the SNR threshold for inclusion in the catalog. It is believed that the addition of prior information (coincidence in redshift as well as position with a likely OC) increases the probability that the detection is real. In fact, all nine of these sources designated as such in the ALFALFA catalog and presented in Table \ref{tab:HIData} were confirmed by the deeper LBW observations. In addition to the catalogued priors, another set of low SNR candidate detections coincident with OCs of know redshift were not included in the published ALFALFA catalog due to polarization mismatch, heavy RFI excision or baseline structure. Thirteen of these ``marginal priors'' were targets in these LBW observations; the chosen targets were selected from a larger set based on careful examination of the HI line and optical data to optimize detection probability. All but one were confirmed despite the relatively poor data quality. As noted in previous ALFALFA papers, we emphasize that the sources categorized as priors should not be used in statistical studies that require completeness, but they can be useful for other studies which do not require strict statistics.

Of particular importance to our aim of increasing the sampling of low mass galaxies in the PPS region is the set of low SNR ALFALFA sources with probable OCs of unknown redshift ("Low SNR, OC, no z"). As noted above, the lack of SDSS spectroscopic coverage in the PPS region gives a higher proportion of such sources in that part of the sky compared to the SDSS spectroscopic footprint. Hence we targeted a large number of such sources particularly those in or near the PPS and in the velocity range c$z <$ 9,000 \kms ~in the observations reported here. In fact, of 295 low SNR sources coincident with possible OCs that were targeted by the LBW observations, 207 (70\%) were confirmed and their ${\rm{H}}{\rm{I}}$ properties are included in Table \ref{tab:HIData}. The 88 that were not observed could simply lie below the detection threshold at their distance (i.e. Figure \ref{fig:span18}) or lie beyond the bandpass limit of 1340 MHz.  Since these were chosen from ALFALFA candidate detections which have been shown to be generally blue and gas rich \citep[\textit{e.g.}][]{2011AJ....142..170H}, it is unlikely that any of them are gas poor. These observations support the expectation that the likelihood of confirmation grows with the SNR of the candidate. It should be noted that observing priority was given to candidates with SNR $>$ 5 so that the lower SNR bins may be underrepresented, but nonetheless, the expected trend of higher reliability at higher SNR is evident.

\begin{figure}[!ht]
\plotone{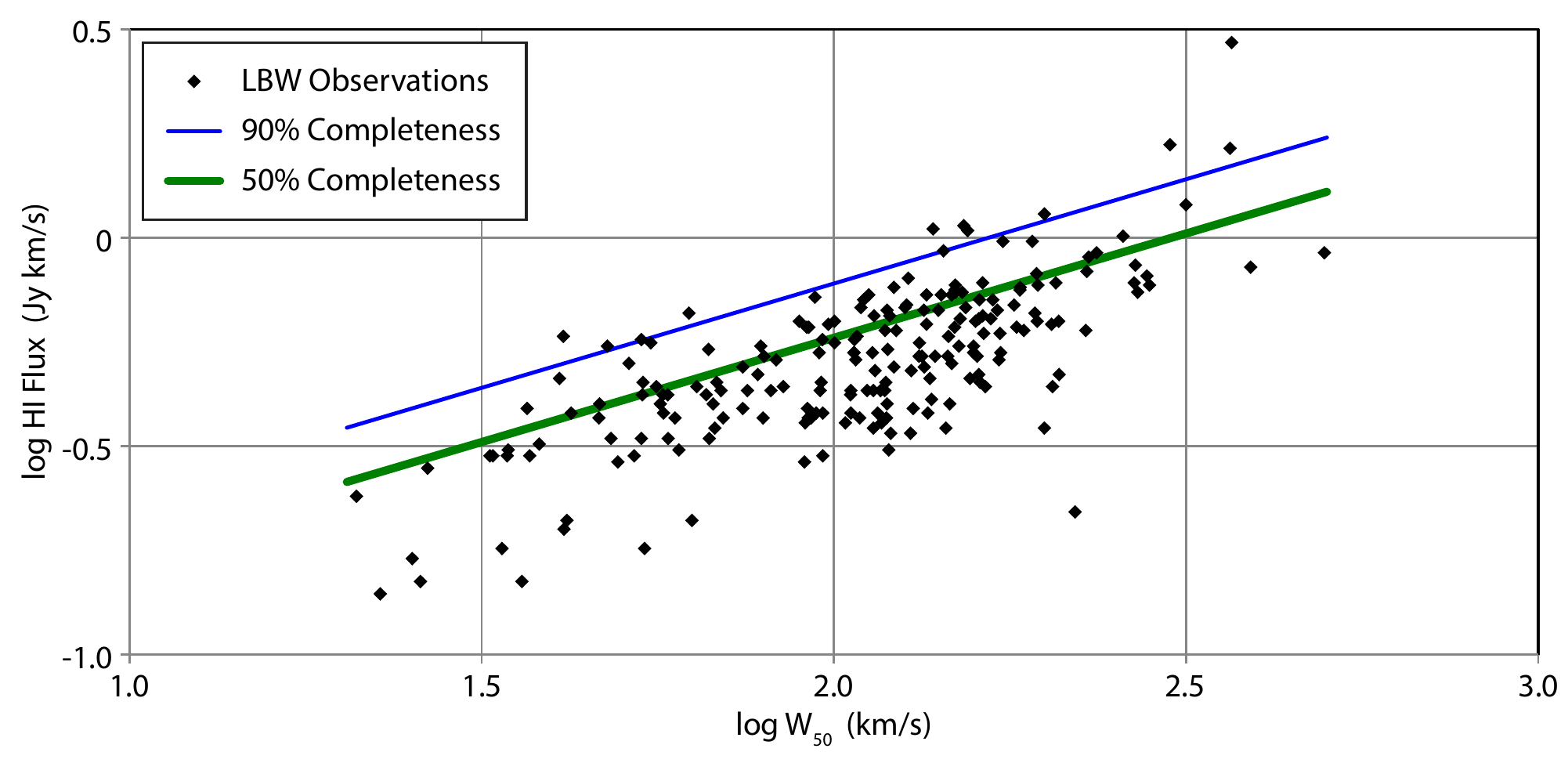}
\caption{Distribution of all low SNR ALFALFA ${\rm{H}}{\rm{I}}$ candidates with possible OCs in logarithmic space depicting the ${\rm{H}}{\rm{I}}$ line width W$_{50}$ and ${\rm{H}}{\rm{I}}$ line flux density as derived from the LBW observations. The thin blue and thick green lines show the 90\% and 50\% completeness relations derived for the ALFALFA high SNR sample in the low velocity width regime log W $<$ 2.5 following equations 6 and 7 of \citet{2011AJ....142..170H}. As evident here, the LBW targets did not meet the detection criteria as high SNR ALFALFA sources but fill the region of lowered ALFALFA completeness. The coincidence of a low SNR ${\rm{H}}{\rm{I}}$ line signal with a low surface brightness star-forming galaxy increases the probability that such a signal is real.
\label{fig:code4complete}}
\end{figure}

As further verification of the utility of these low SNR ALFALFA candidates with possible OCs, Figure \ref{fig:code4complete} shows the relationship of the ${\rm{H}}{\rm{I}}$ profile velocity width W$_{50}$ versus the ${\rm{H}}{\rm{I}}$ line flux density, on logarithmic scales, for the detected low SNR galaxies with OCs as derived from their LBW ${\rm{H}}{\rm{I}}$ line spectra. Also plotted are the ALFALFA completeness relations determined at the 90\% (black line) and 50\% (red line) for the single range of log W$_{50} <$ 2.5, as given in Equations 6 and 7 of \citet{2011AJ....142..170H}. As is evident in the figure, nearly all of the sources did not meet the criteria for inclusion in the ALFALFA catalog, but probe the range of low ALFALFA completeness. 

\subsection{Optically-Selected Targets} \label{subsec:SDSStargets}

Although a robust statistical analysis is not possible at this point because of the heterogeneity of the sample, we also selected a set of candidates suggested by their optical appearance to likely be gas rich and detectable in ${\rm{H}}{\rm{I}}$ at the distance of PPS, between 4,000 \kms~and 7,000 \kms. Selection criteria focused on optical morphology (spiral and/or diffuse appearance), low surface brightness and blue color, similar to the majority of typical low mass, star-forming galaxies found in the ALFALFA catalog. An analysis of the typical optical properties of the ALFALFA population was presented by \citet{2012ApJ...756..113H}. Following the scaling relations used in that work, SQL queries of the SDSS database can identify galaxies showing similar optical properties as the lower HI mass galaxies detected by ALFALFA. For the purpose of the LBW observations conducted here, we added targets selected from the SDSS based on color, surface brightness and their optical morphology. In a few cases, there was a known redshift, placing the target in the PPS range. As evident in Table \ref{tab:CategoryRates}, galaxies identified by their optical characteristics were detected at a significant rate (62.5\%). While the optical selection criteria for the targets reported here were not uniform or statistically robust, our pilot experiment reinforced the potential to use optical photometry alone to identify additional PPS galaxies which are below the ALFALFA sensitivity limit but which are detected by short, targeted LBW observations. This approach has been further extended and serves as the basis for target selection for the APPSS to be presented in future works.

\section{Conclusions} \label{sec:Conclusions}

The results of these three Arecibo LBW observing programs in the direction of PPS have tested the reliability of detections made as part of the ALFALFA blind ${\rm{H}}{\rm{I}}$ survey:

\begin{itemize}

\item{
A small percentage of high SNR ALFALFA sources are spurious either due to random noise fluctuations in such a large dataset or residual RFI contamination. Our result is consistent with the simulations of reliability done by \citet{2007AJ....133.2087S}.
}

\item{
At the same time, the program has shown that, at least in the PPS region, the large majority of the high SNR galaxies without an identified OC are real.  Some are tidal remnants and one (AGC310858) has been identified by \citet{2017ApJ...842..133L} to be an HI-bearing ultra diffuse galaxy.  
}

\item{
Among the lowest redshift (c$z <$ 1,000 \kms) low SNR ALFALFA targets, those lacking an OC are virtually all spurious since we detected zero of 84.  A detection of one of these sources, a truly ``dark dwarf'', would have been very interesting and the lack of such a detection leaves their existence as inconclusive.
}

\item{ In contrast to the previous conclusion, the
low SNR ALFALFA sources which appear to be associated with blue, star-forming OCs are very likely to be real; we detected 70\% of those that were targeted.
}
\end{itemize}

Further, we have shown the efficacy of using pointed LBW observations to detect lower ${\rm{H}}{\rm{I}}$ mass galaxies as shown in Figure \ref{fig:span18}. Adopting selection criteria based on optical morphology, low surface brightness and blue color allows a way to sample additional low ${\rm{H}}{\rm{I}}$ mass galaxies in the fixed volume in and around PPS by probing the ${\rm{H}}{\rm{I}}$ mass function below the detection threshold of ALFALFA. The demonstrated success of this pilot program to select targets through SQL queries based on SDSS photometric criteria establishes confidence to utilize this method for selecting targets for the UAT APPSS for which observations are ongoing.

\section{Acknowledgements} \label{sec:Acknowledgements}

With great sadness, we acknowledge the contribution to the observational program reported here
of our late colleague and dear friend Ron Olowin. 

We are grateful for the contributions of all members of the Undergraduate ALFALFA Team.  This work has been supported by NSF grants AST-1211005, AST-1637339 and AST-1637271. RG, MPH, and LL
and MGJ acknowledge support from  NSF grants AST-1107390 and AST-1714828 and by grants from the Brinson Foundation. MGJ acknowledges support from the grant 
AYA2015-65973-C3-1-R (MINECO/FEDER, UE).

The Arecibo Observatory has been operated by SRI International under a cooperative agreement with the National Science Foundation (AST-1100968), and in alliance with Ana G. M\'endez-Universidad Metropolitana, and the Universities Space Research Association.

Funding for the SDSS and SDSS-II has been provided by the Alfred P. Sloan Foundation, the participating institutions, the National Science Foundation, the US Department of Energy, the NASA, the Japanese Monbukagakusho, the Max Planck Society, and the Higher Education Funding Council for England. The SDSS Web site is http://www.sdss.org/. The SDSS is managed by the Astrophysical Research Consortium for the participating institutions. 

\facilities{Arecibo}
\software{IDL}

\end{document}